\documentclass[dvips,12pt,a4paper]{article}
\usepackage{epsfig}
\usepackage{amsmath}

\def\lsim{\mathrel{\rlap{\raise 2.5pt \hbox{$<$}}\lower 2.5pt
\hbox{$\sim$}}}
\def\gsim{\mathrel{\rlap{\raise 2.5pt \hbox{$>$}}\lower 2.5pt
\hbox{$\sim$}}}

\def\thW{\theta_{\rm W}}
\def\GeV{{\rm GeV}}
\def\TeV{{\rm TeV}}
\def\dd{{\rm d}}
\def \sup{^{\vphantom{2}}}

\begin{document}

\thispagestyle{empty}
\phantom{anchor}
\vspace*{-40mm}

\begin{flushright}
 {\tt University of Bergen, Department of Physics}    \\[1mm]
 {\tt Scientific/Technical Report No.1999-01}    \\[1mm]
 {\tt ISSN 0803-2696} \\[1mm]
 {hep-ph/9902270} \\[1mm]
 {February 1999}   \\
\end{flushright}
\vspace*{12mm}

\begin{center}
{\bf{\Large Measuring Trilinear Higgs Couplings in the 
MSSM}}\footnote{Presented at 
{\sl VIIIth UNESCO St.~Petersburg International School of Physics},
May 25 -- June 4, 1998.
To be published in the Proceedings}
\vskip 0.5cm
P. Osland$^{a}$ and P. N. Pandita$^{b}$\\
$^a$ Department of Physics, University of Bergen, 
N-5007 Bergen, Norway\\
$^b$ Department of Physics, North Eastern Hill University,
Shillong 793 022, India
\end{center}

\begin{abstract}
Trilinear couplings of the neutral $CP$-even
Higgs bosons in the Minimal Supersymmetric 
Standard Model (MSSM) can be measured through
the multiple production of the lightest $CP$-even Higgs boson ($h$)
at  high-energy $e^+  e^-$ colliders. This includes 
the production of 
the heavier $CP$-even Higgs boson ($H$) via
$e^+e^- \rightarrow ZH$, in association with 
the $CP$-odd Higgs boson ($A$) in $e^+e^- \rightarrow AH$, 
or via $e^+e^- \rightarrow \nu_e \bar\nu_e H$, 
with $H$ subsequently decaying through $H \rightarrow hh$.
These processes can enable one to measure the trilinear Higgs couplings 
$\lambda_{Hhh}$ and
$\lambda_{hhh}$, which can be used to theo\-retically reconstruct the
Higgs potential. 
We delineate the regions of the MSSM parameter space in which these
trilinear Higgs couplings could be measured. 
\end{abstract}

\section{Introduction}
Supersymmetry is at present the only known
framework in which the Higgs
sector~\cite{GHKD} of the Standard Model (SM),
so crucial for its internal consistency, is natural~\cite{HPN}.
The minimal version of the Supersymmetric 
Standard Model (MSSM) contains two Higgs doublets $(H_1, H_2)$ 
with opposite hypercharges: $Y(H_1) = -1$, $Y(H_2) = +1$, so as to 
generate masses for up- and down-type
quarks (and leptons), and to cancel gauge anomalies. 
After spontaneous symmetry breaking induced by the neutral 
components of $H_1$ and $H_2$ obtaining vacuum 
expectation values, $\langle H_1\rangle = v_1$, 
$\langle H_2\rangle = v_2$, $\tan\beta = v_2/v_1$, 
the MSSM contains two neutral $CP$-even ($h$, $H$), one neutral 
$CP$-odd ($A$), and two charged ($H^{\pm}$) Higgs bosons \cite{GHKD}. 
Because of gauge invariance and supersymmetry,
all the Higgs masses
and the Higgs couplings in the MSSM can be described (at tree level) 
in terms of only  two parameters, which are usually chosen to be
$\tan\beta$ and $m_A$, the mass of the $CP$-odd Higgs boson.

In particular, all the trilinear self-couplings of the physical
Higgs particles can be predicted theoretically (at the tree level)
in terms of $m_A$ and $\tan\beta$. Once a light Higgs boson is 
discovered, the measurement of these trilinear couplings can be used to 
reconstruct the Higgs potential of the MSSM. This will go a long way
toward establishing the Higgs mechanism as the basic mechanism
of spontaneous symmetry breaking in gauge theories. Although the 
measurement of all the Higgs couplings in the MSSM is a difficult task,
preliminary theoretical investigations by Plehn, Spira and Zerwas
\cite{PSZ}, and by Djouadi, Haber and Zerwas (DHZ) \cite{DHZ}, 
of the measurement of these couplings at the LHC and
at a high-energy $e^+ e^-$ linear collider, respectively, are encouraging.

We have considered in  detail \cite{OP98}
the question of possible measurements
of the trilinear Higgs couplings of the MSSM at a high-energy $e^+ e^-$ 
linear collider. We assume that such a facility will operate at
an energy of 500~GeV with an integrated luminosity per year of  
${\mathcal L}_{\rm int} = 500~\mbox{fb}^{-1}$ \cite{NLC}.
(This is a factor of 10 more than the earlier estimate.)
In a later phase one may envisage an upgrade to an energy of 1.5~TeV.

The trilinear Higgs couplings that are of interest are 
$\lambda_{Hhh}$, $\lambda_{hhh}$, and $\lambda_{hAA}$, 
involving both the $CP$-even and $CP$-odd Higgs bosons.
The couplings $\lambda_{Hhh}$ and $\lambda_{hhh}$
are rather small with respect to the corresponding trilinear coupling
$\lambda_{hhh}^{\rm SM}$ in the SM (for a given mass 
of the lightest Higgs boson $m_h$), 
unless $m_h$ is close to the upper value (decoupling limit).
The coupling $\lambda_{hAA}$ remains small for all parameters.

Throughout, we include one-loop radiative corrections
\cite{ERZ1} to the Higgs sector in the effective potential
approximation. In particular, we take into account 
the parameters $A$ and $\mu$, the soft supersymmetry
breaking trilinear parameter and the bilinear Higgs(ino)  
parameter in the superpotential, respectively, and as a consequence
the left--right mixing in the squark sector, in our calculations. 
We thus include all the relevant parameters of the MSSM in our study
\cite{OP98}.
Related work has recently been presented by Dubinin and Semenov 
\cite{DubSem}.

For a given value of $m_h$, the values of these
couplings significantly depend on the soft supersymmetry-breaking 
trilinear parameter $A$, as well as on $\mu$, 
and thus on the resulting mixing in the squark sector.
Since the trilinear couplings tend to be small,
and depend on several parameters, their effects are somewhat difficult
to estimate.

The dominant source of multiple production
of the Higgs ($h$) boson, is through Higgs-strahlung of $H$, and 
through production of $H$ in association with the $CP$-odd Higgs 
boson.  This source of multiple production can be used to 
extract the trilinear Higgs coupling $\lambda_{Hhh}$.
The non-resonant fusion mechanism for multiple
$h$ production, $e^+e^-\to \nu_e\bar\nu_e hh$, involves
two trilinear Higgs couplings, $\lambda_{Hhh}$ and $\lambda_{hhh}$,
and is useful for extracting $\lambda_{hhh}$.

\section{The Higgs Sector of the MSSM}

At the tree level, the Higgs sector of the MSSM 
is described by two parameters, which can be conveniently chosen as
$m_A$ and $\tan\beta$ \cite{GHKD}. There are, however, substantial
radiative corrections to the $CP$-even neutral Higgs masses and
couplings~\cite{ERZ1,ERZ2}. 
They are, in general, positive,
and they shift the mass of the lightest MSSM Higgs boson upwards.

The Higgs mass falls rapidly at small values of $\tan\beta$.
Since the LEP experiments are obtaining 
lower bounds on the mass of the lightest Higgs boson, they are
beginning to rule out significant parts of the small-$\tan\beta$ 
parameter space, depending on the model assumptions.
ALEPH finds a lower limit of $m_h>72.2$~GeV, irrespective of $\tan\beta$,
and a limit of $\sim 88$~GeV for $1<\tan\beta\lsim2$ \cite{ALEPH98}.
We take $\tan\beta=2$ to be a representative value.

\setcounter{equation}{0}
\section{Trilinear Higgs couplings}

In units of $gm_Z/(2\cos\thW)=(\sqrt{2}G_F)^{1/2}m_Z^2$,
the relevant tree-level trilinear Higgs couplings are given by
\begin{eqnarray} 
\lambda_{Hhh}^0 & = & 2\sin2\alpha \sin(\beta + \alpha) - \cos 2\alpha
\cos(\beta + \alpha), \\
\label{Eq:lambda-Hhh0}
\lambda_{hhh}^0 & = & 3 \cos2\alpha \sin(\beta + \alpha), \\
\label{Eq:lambda-hhh0}
\lambda_{hAA}^0 & = & \cos2\beta \sin(\beta + \alpha),      
\label{Eq:lambda-hAA0}
\end{eqnarray}
with $\alpha$ the mixing angle in the $CP$-even Higgs sector, which can
be calculated in terms of the parameters appearing in the $CP$-even
Higgs mass matrix. 
The dominant one-loop radiative corrections are proportional
to $(m_t/m_W)^4$ \cite{BBSP}.

\begin{figure}[htb]
\refstepcounter{figure}
\label{Fig:lam-mh}
\addtocounter{figure}{-1}
\begin{center}
\setlength{\unitlength}{1cm}
\begin{picture}(15,8.5)
\put(-1.5,1)
{\mbox{\epsfysize=8.5cm\epsffile{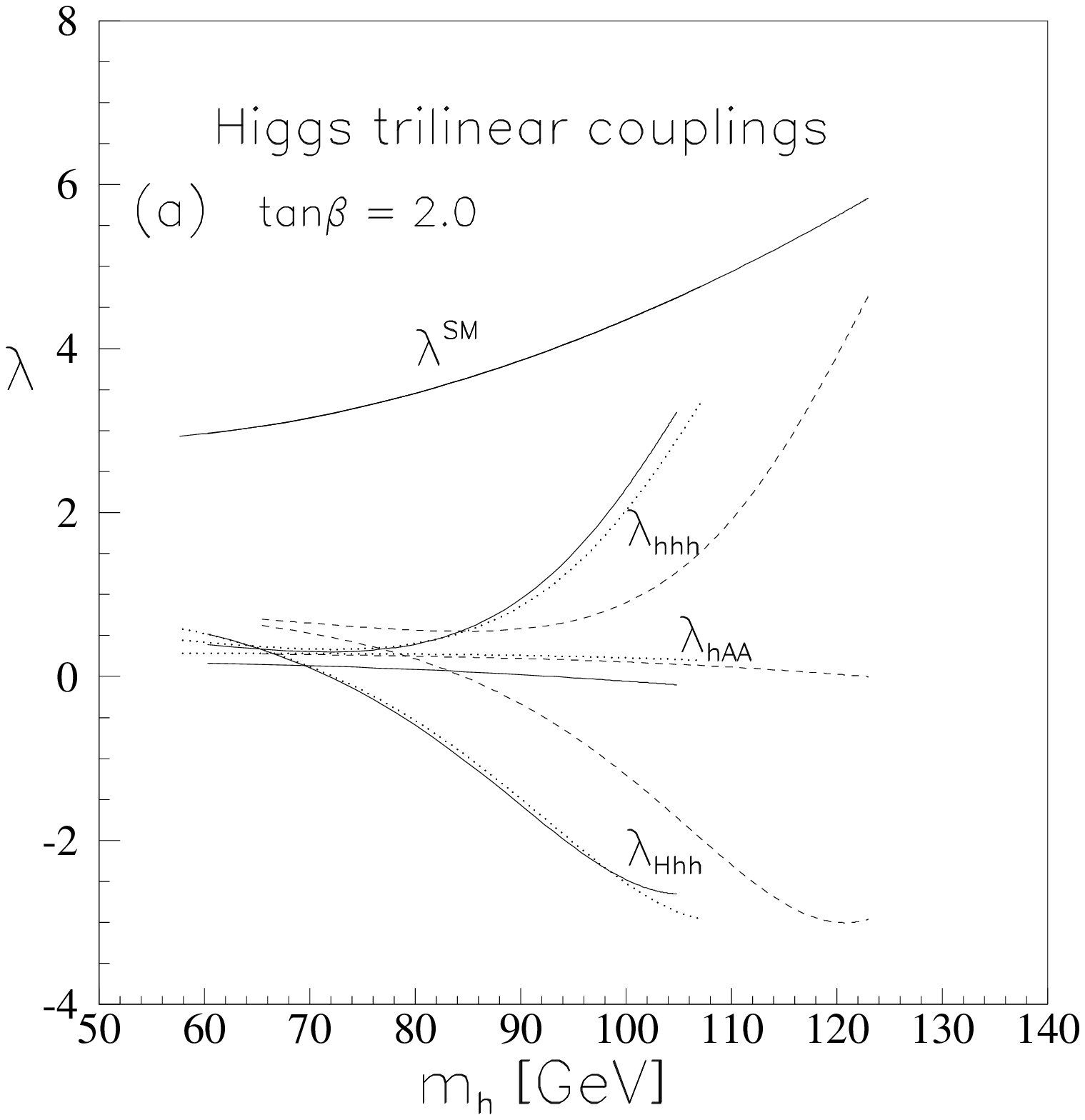}}
 \mbox{\epsfysize=8.5cm\epsffile{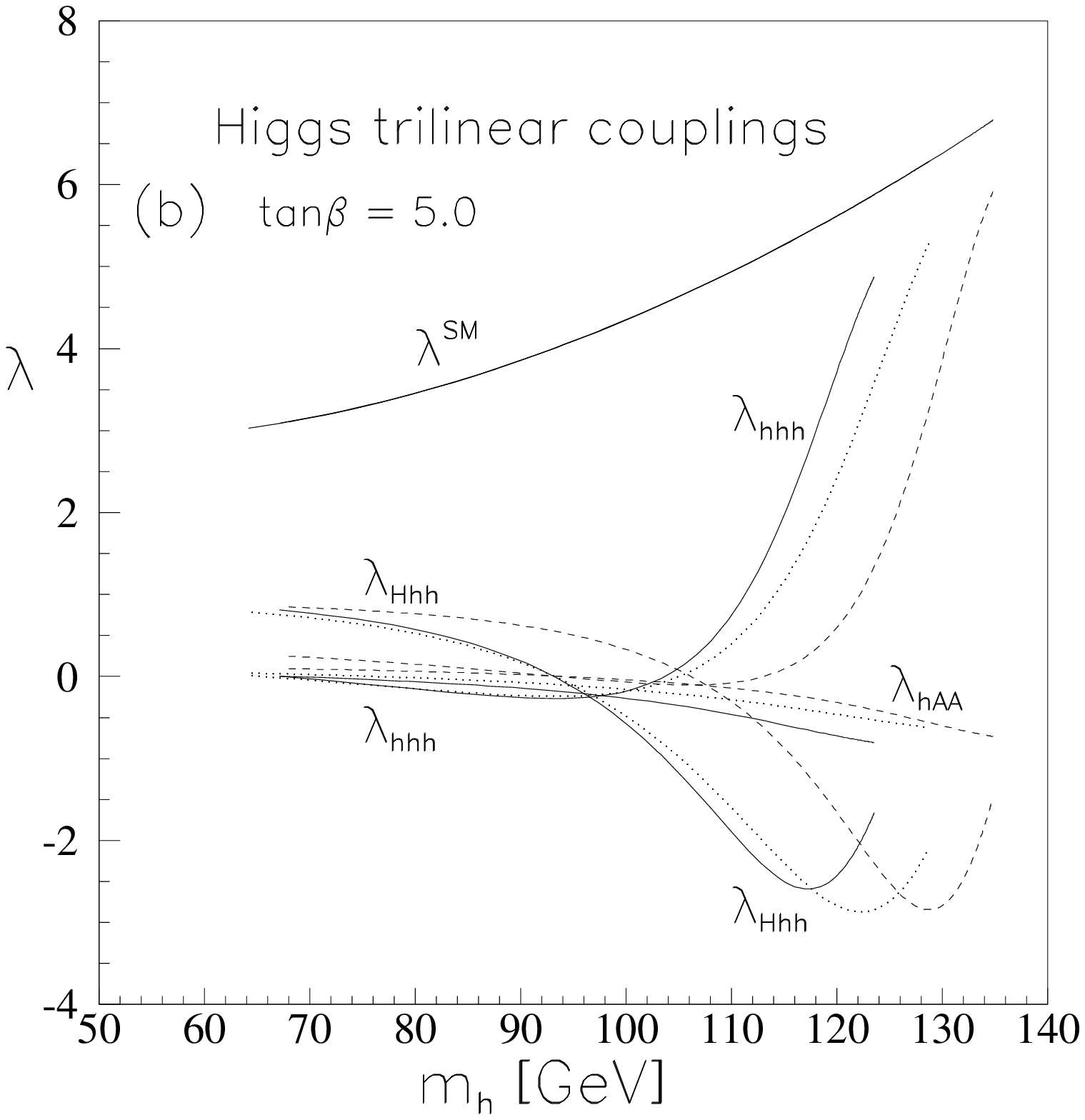}}}
\end{picture}
\vspace*{-8mm}
\caption{Trilinear Higgs couplings $\lambda_{Hhh}$, $\lambda_{hhh}$ and
$\lambda_{hAA}$ as functions of $m_h$ for two values of $\tan\beta$:
(a) $\tan\beta=2.0$,
(b) $\tan\beta=5.0$.
Each coupling is shown for three cases of the
mixing parameters:
no mixing ($A=0$, $\mu=0$, solid),
mixing with $A=1$~TeV and $\mu=-1$~TeV (dotted),
as well as 
$A=1$~TeV and $\mu=1$~TeV (dashed).}
\end{center}
\end{figure}

The trilinear couplings depend significantly on $m_A$,
and thus also on $m_h$. This is shown in Fig.~\ref{Fig:lam-mh},
where we compare $\lambda_{Hhh}$, $\lambda_{hhh}$ and $\lambda_{hAA}$
for three different values of $\tan\beta$,
and the SM quartic coupling $\lambda^{\rm SM}$ (which also includes
one-loop radiative corrections \cite{SirZuc}).

At low values of $m_h$, the MSSM trilinear couplings are rather small.
For some value of $m_h$ the couplings $\lambda_{Hhh}$ and $\lambda_{hhh}$
start to increase in magnitude, whereas $\lambda_{hAA}$ remains small.
The values of $m_h$ at which they start becoming significant
depend crucially on $\tan\beta$.
For $\tan\beta=2$ (Fig.~\ref{Fig:lam-mh}a) this transition
takes place around $m_h\sim 90$--100~GeV, whereas
for $\tan\beta=5$ the critical value of $m_h$ increases
to 100--110 (see Fig.~\ref{Fig:lam-mh}b).
In this region, the actual values of $\lambda_{Hhh}$ and $\lambda_{hhh}$
(for a given value of $m_h$) change significantly if $A$ becomes
large and positive. 
A non-vanishing squark-mixing parameter $A$ is thus quite 
important. Also, for special values of the parameters,
the couplings may vanish \cite{DKZ1}.

To sum up the behaviour of the trilinear couplings, we note that
$\lambda_{Hhh}$ and  $\lambda_{hhh}$ are small for 
$m_h \lsim 100$--120~GeV, depending on the value of $\tan\beta$. 
However, as $m_h$ approaches its maximum value, 
which is reached rapidly as $m_A$ becomes large,
$m_A \gsim 200$~GeV, these trilinear couplings become large.
Thus, as functions of $m_A$, the trilinear couplings
$\lambda_{Hhh}$ and $\lambda_{hhh}$ are large
for most of the parameter space.
We also note that, for large values of $\tan\beta$, $\lambda_{Hhh}$
tends to be relatively small, whereas $\lambda_{hhh}$ becomes large,
if also $m_A$ (or, equivalently, $m_h$) is large.

\setcounter{equation}{0}
\section{Production mechanisms}

The different mechanisms for the multiple production of the MSSM 
Higgs bosons in $e^+ e^-$ collisions have been discussed by DHZ.
The dominant mechanism for the production of multiple  
$CP$-even light Higgs bosons ($h$) is through the 
production of the heavy $CP$-even Higgs boson $H$, which then decays
via $H \rightarrow hh$. The heavy Higgs boson $H$ can be produced
by $H$-strahlung, in association with $A$, 
and by the resonant $WW$ fusion mechanism. These mechanisms
for multiple production of $h$
\begin{eqnarray}
\left. \begin{array}{ccc}
e^+e^- & \rightarrow & ZH,AH \\ 
e^+e^- & \rightarrow & \nu_e \bar \nu_e H
\end{array}
\right\}, \qquad H \rightarrow hh, \label{Eq:res-Hhh} 
\end{eqnarray}
are shown in Fig.~\ref{Fig:Feynman-resonant}. 
All the diagrams of Fig.~\ref{Fig:Feynman-resonant} involve the
trilinear coupling $\lambda_{Hhh}$.

\begin{figure}[htb]
\refstepcounter{figure}
\label{Fig:Feynman-resonant}
\addtocounter{figure}{-1}
\begin{center}
\setlength{\unitlength}{1cm}
\begin{picture}(16,7)
\put(2,-12)
{\mbox{\epsfxsize=16cm\epsffile{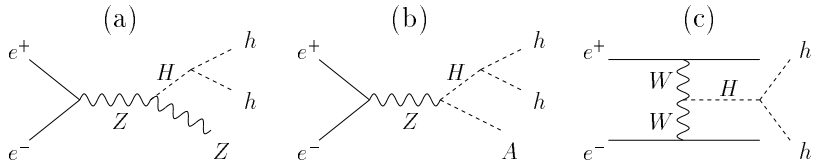}}}
\end{picture}
\vspace*{-50mm}
\caption{Feynman diagrams for the resonant production
of $hh$ final states in $e^+ e^-$ collisions.}
\end{center}
\end{figure}
\vspace*{-5mm}


A background to (\ref{Eq:res-Hhh}) comes from the production of the 
pseudoscalar $A$ in association with $h$ and its subsequent decay to $hZ$
\begin{equation}
e^+e^- \rightarrow hA, \qquad A \rightarrow hZ, \label{Eq:bck-hA}
\end{equation}
leading to $Zhh$ final states.
A second mechanism for $hh$ production is double Higgs-strahlung 
in the continuum with a $Z$ boson in the final state,
\begin{equation}
e^+e^-  \rightarrow Z^* \rightarrow Zhh. \label{Eq:Zstar}
\end{equation} 
We note that the non-resonant analogue of the Feynman diagram of 
Fig.~\ref{Fig:Feynman-resonant}b
involves, apart from the coupling $\lambda_{Hhh}$,  
the trilinear Higgs coupling $\lambda_{hhh}$ as well.

\begin{figure}[htb]
\refstepcounter{figure}
\label{Fig:Feynman-nonres-WW}
\addtocounter{figure}{-1}
\begin{center}
\setlength{\unitlength}{1cm}
\begin{picture}(16,7)
\put(1,-12)
{\mbox{\epsfxsize=16cm\epsffile{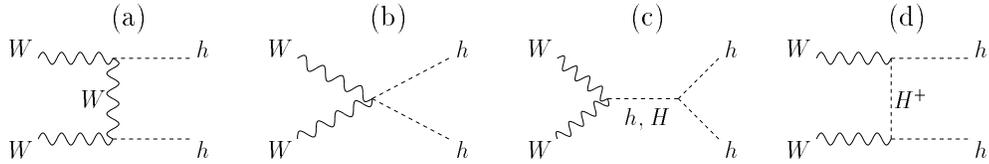}}}
\end{picture}
\vspace*{-50mm}
\caption{Feynman diagrams for the non-resonant $WW$ fusion
mechanism for the production of $hh$ states in $e^+ e^-$ collisions.}
\end{center}
\end{figure}
\vspace*{-5mm}

Finally, there is a mechanism of multiple production of the lightest Higgs
boson through non-resonant $WW$ fusion in the continuum
(see Section~7):
\begin{equation}
e^+e^-  \rightarrow \bar \nu_e \nu_e W^* W^* \rightarrow 
\bar \nu_e \nu_e hh. \label{Eq:WW-fusion}
\end{equation}

It is important to note that all the diagrams of  
Fig.~\ref{Fig:Feynman-resonant}
involve the trilinear coupling $\lambda_{Hhh}$ only. On the other hand,
the non-resonant analogue of Fig.~\ref{Fig:Feynman-resonant}b, 
and  Fig.~\ref{Fig:Feynman-nonres-WW}c involve both
the trilinear Higgs couplings $\lambda_{Hhh}$ and $\lambda_{hhh}$.

\setcounter{equation}{0}
\section{Higgs-strahlung and Associated Production of $H$}
The dominant source for the production of
multiple Higgs bosons ($h$) in $e^+ e^-$ collisions is through 
the production of the heavier $CP$-even Higgs boson $H$ either via
Higgs-strahlung or in association with $A$,  
followed, if kinematically allowed, by the cascade decay 
$H \rightarrow hh$.
The cross sections for these processes can be found in
\cite{PocZsi,GETAL}.

\begin{figure}[htb]
\refstepcounter{figure}
\label{Fig:sigma-500-1500}
\addtocounter{figure}{-1}
\begin{center}
\setlength{\unitlength}{1cm}
\begin{picture}(15,8.5)
\put(-1.5,1)
{\mbox{\epsfysize=8.5cm\epsffile{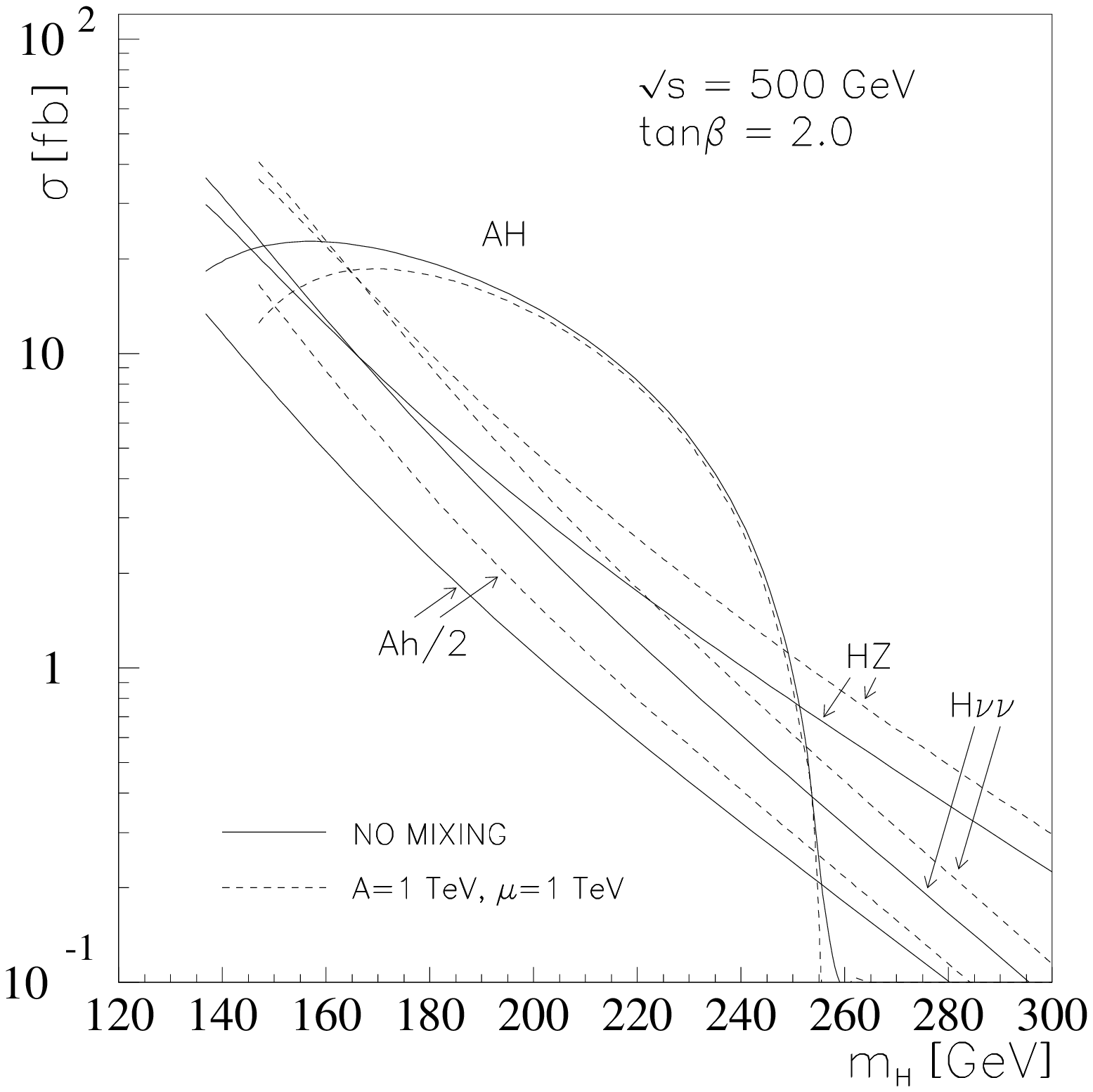}}
 \mbox{\epsfysize=8.5cm\epsffile{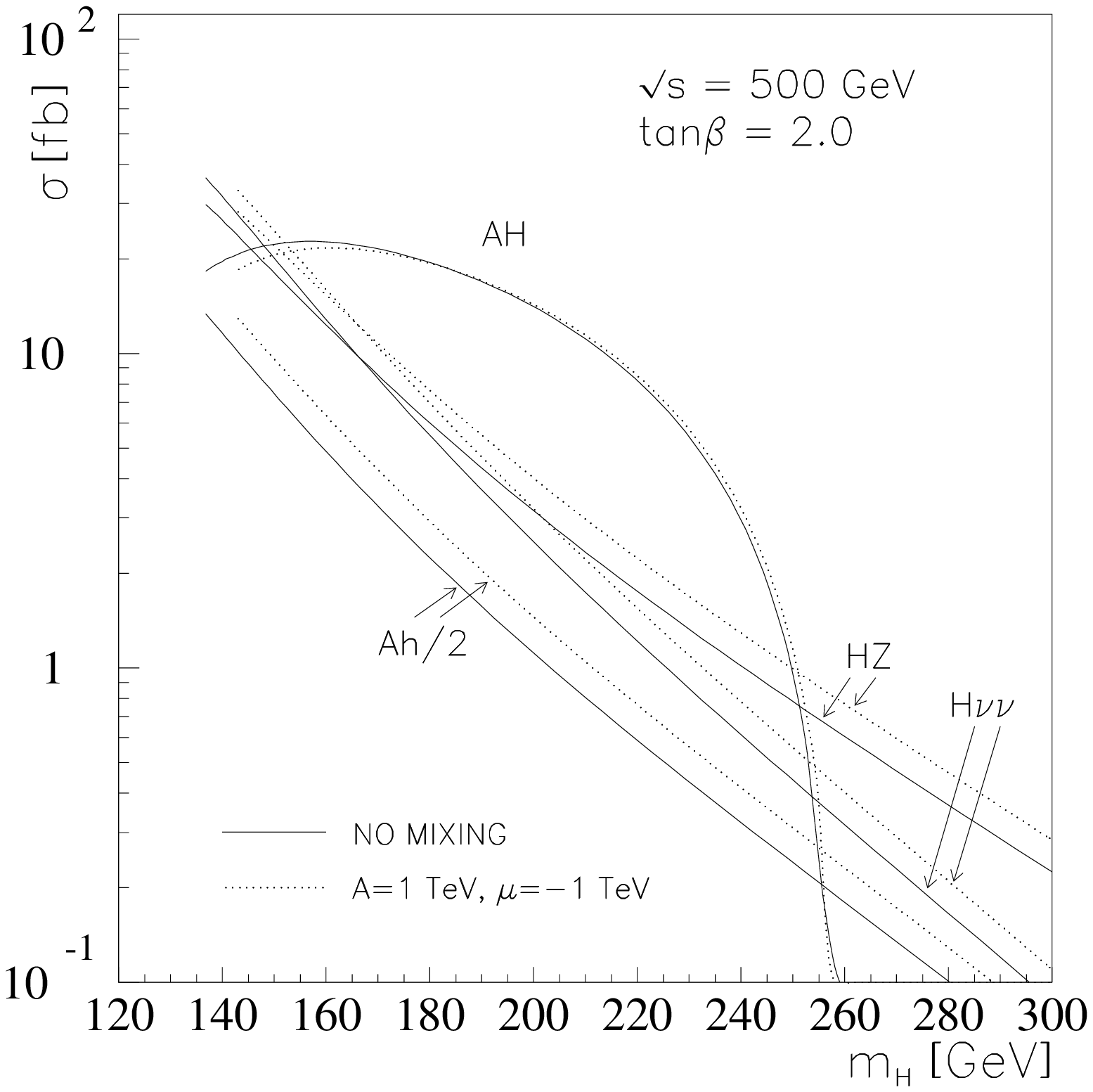}}}
\end{picture}
\vspace*{-8mm}
\caption{Cross sections for the production of the heavy Higgs
boson $H$ in $e^+ e^-$ collisions,
and for the background process in which $Ah$ is produced. 
Solid curves are for no mixing, $A=0$, $\mu=0$. 
Dashed and dotted curves refer to mixing, as indicated.}
\end{center}
\end{figure}

In Fig.~\ref{Fig:sigma-500-1500} we plot the cross sections 
for the 
$e^+e^-$ centre-of-mass energies $\sqrt s = 500~\GeV$,
as functions of the Higgs mass $m_H$ and for $\tan\beta = 2.0$. 
For large values of the mass $m_A$ of the pseudoscalar 
Higgs boson, all the Higgs bosons, 
except the lightest one ($h$), become heavy and  decouple~\cite{HABER1} 
from the rest of the spectrum. 

At values of $\tan\beta$ that are not too large, the trilinear
$Hhh$ coupling $\lambda_{Hhh}$ can be measured by the decay process
$H \rightarrow hh$, which has a width proportional to $\lambda_{Hhh}^2$.
However, this is possible only if the decay is kinematically
allowed, and the branching ratio is sizeable.
In Fig.~\ref{Fig:BR-H-A} we show the branching ratios (at $\tan\beta=2$)
for the main decay modes of the heavy $CP$-even Higgs boson 
as a function of the $H$ mass.
Apart from the $hh$ decay mode, the other important decay modes 
are $H \rightarrow WW^*$, $ZZ^*$. 
We note that the couplings of $H$ 
to gauge bosons can be measured through the production cross sections
for $e^+e^- \rightarrow \nu_e\bar\nu_eH$; therefore the branching
ratio $BR(H \rightarrow hh)$ can be used to measure the triple Higgs
coupling $\lambda_{Hhh}$. 

\begin{figure}[htb]
\refstepcounter{figure}
\label{Fig:BR-H-A}
\addtocounter{figure}{-1}
\begin{center}
\setlength{\unitlength}{1cm}
\begin{picture}(15,8.5)
\put(-1.5,1)
{\mbox{\epsfysize=8.5cm\epsffile{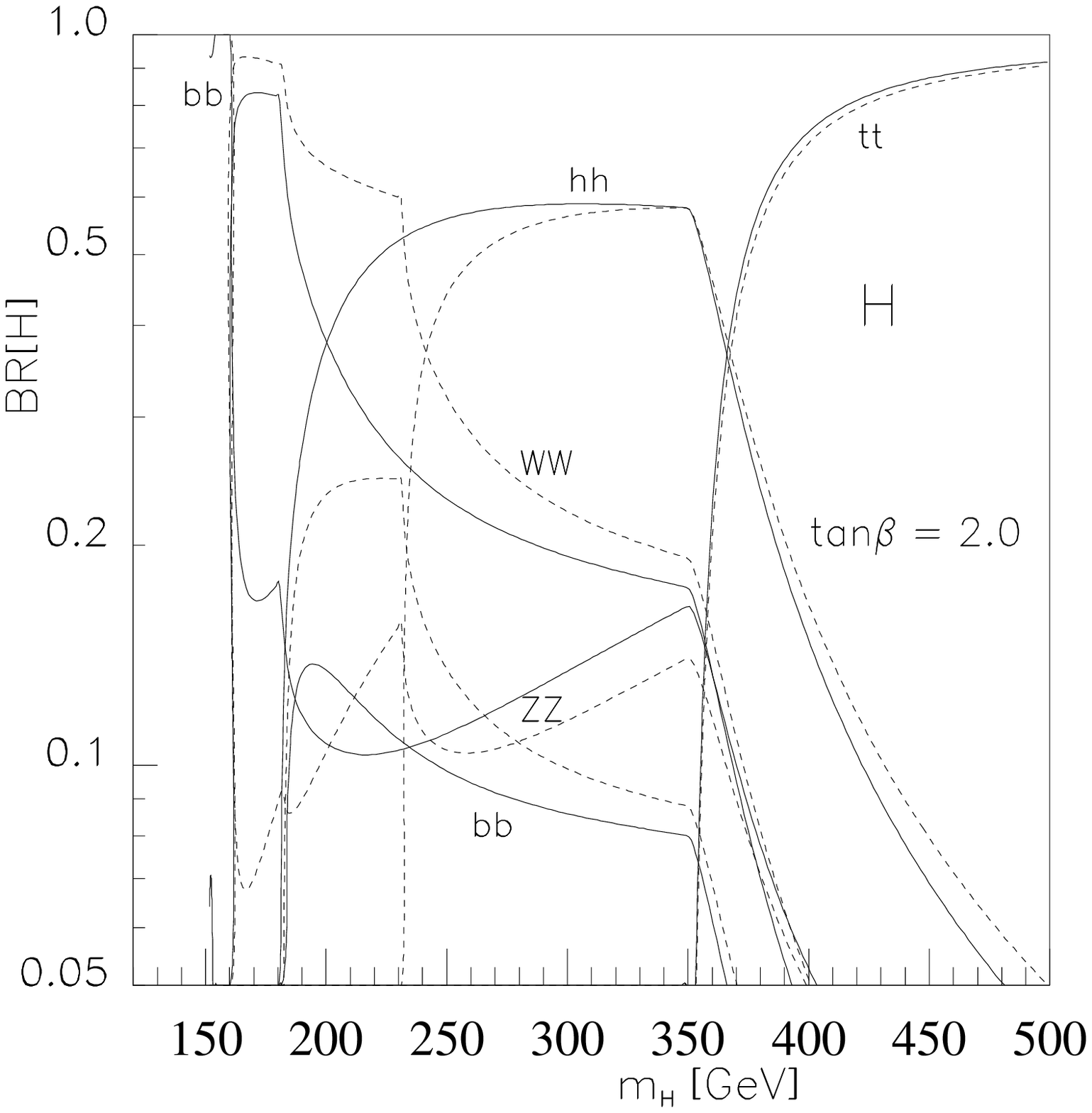}}
 \mbox{\epsfysize=8.5cm\epsffile{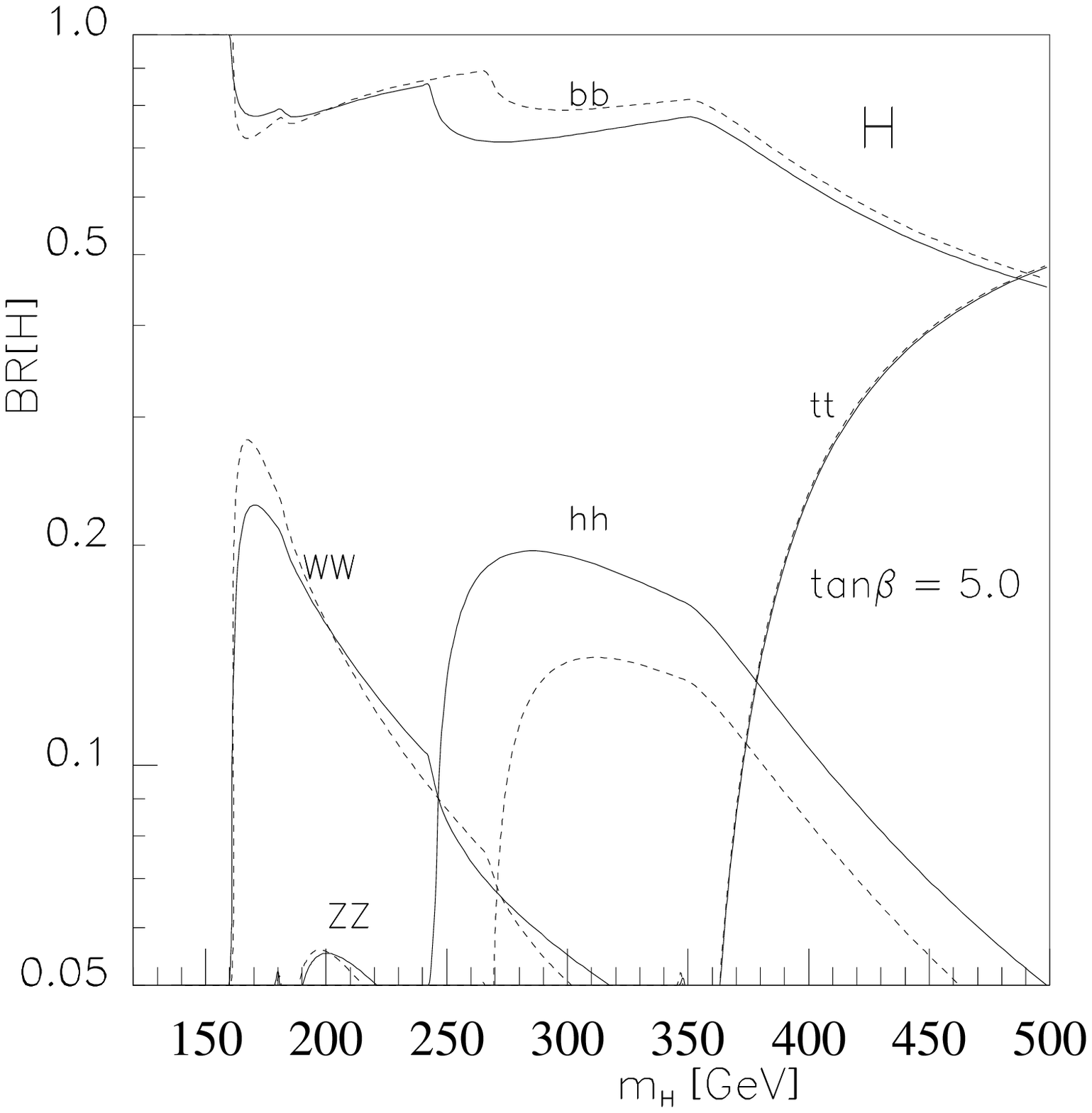}}}
\end{picture}
\vspace*{-8mm}
\caption{Branching ratios for the decay modes of the $CP$-even
heavy Higgs boson $H$, for $\tan\beta = 2.0$ and 5.0.
Solid curves are for no mixing, $A=0$, $\mu=0$. 
Dashed and dotted curves refer to mixing, as indicated.}
\end{center}
\end{figure}

For increasing values of $\tan\beta$, the $Hhh$ coupling gradually 
gets weaker (see Fig.~\ref{Fig:lam-mh}),
and hence the prospects for measuring $\lambda_{Hhh}$ diminish.
This is also indicated in Fig.~\ref{Fig:BR-H-A}, where we show the
$H$ branching ratios for $\tan\beta=5$.

\begin{figure}[htb]
\refstepcounter{figure}
\label{Fig:hole}
\addtocounter{figure}{-1}
\begin{center}
\setlength{\unitlength}{1cm}
\begin{picture}(15,8.5)
\put(-1.5,1)
{\mbox{\epsfysize=8.5cm\epsffile{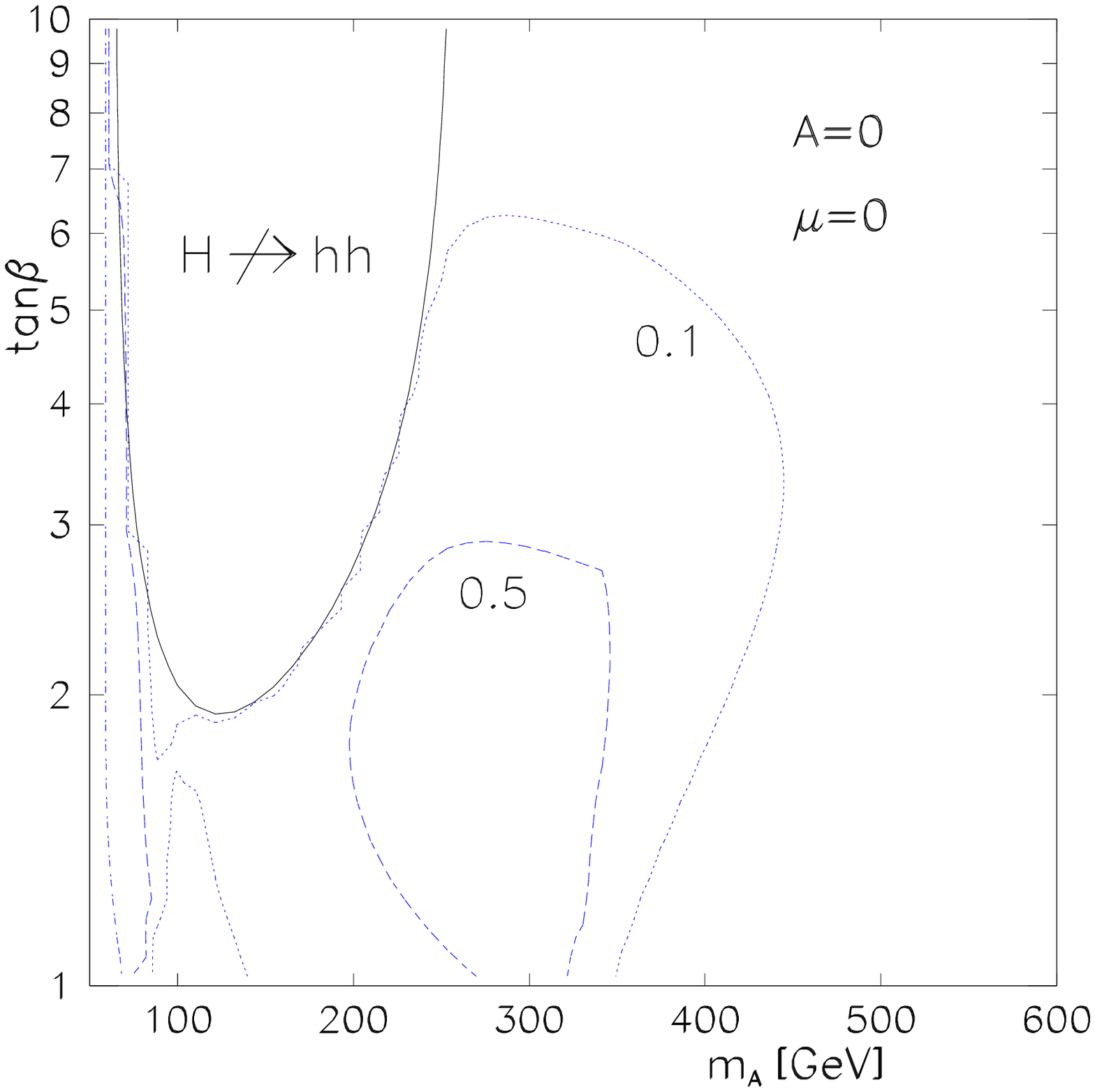}}
 \mbox{\epsfysize=8.5cm\epsffile{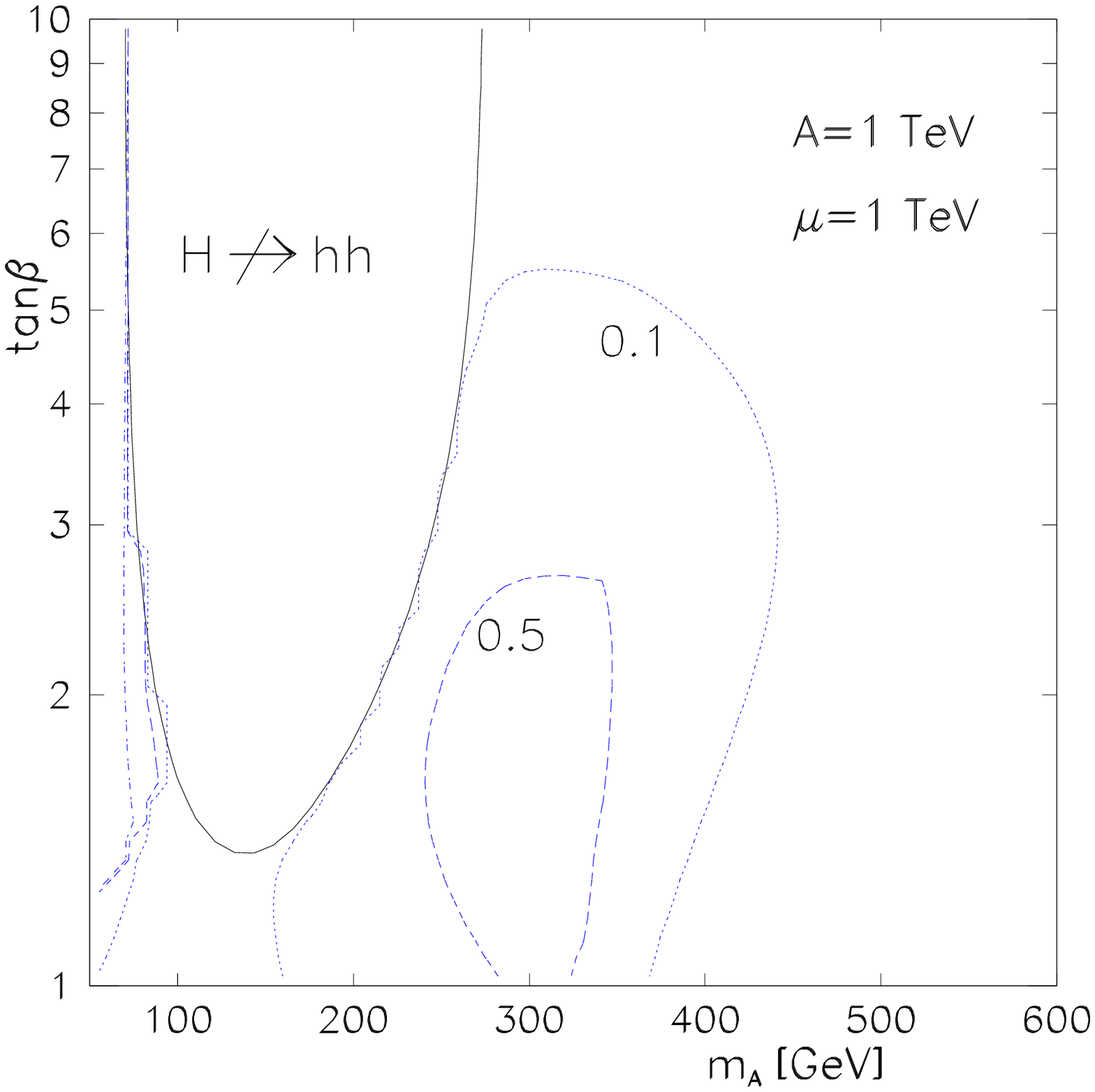}}}
\end{picture}
\vspace*{-8mm}
\caption{The region in the $m_A$--$\tan\beta$ plane where the decay
$H\to hh$ is kinematically {\em forbidden} is indicated by a solid 
line contour.
Also given are contours at which the branching ratio equals 0.1 (dotted),
0.5 (dashed) and 0.9 (dash-dotted, at the far left).}
\end{center}
\end{figure}

There is actually a sizeable region in the $m_A$--$\tan\beta$ plane
where the decay $H\to hh$ is kinematically forbidden.
This is shown in Fig.~\ref{Fig:hole}, where
we also display the regions where the $H\to hh$
branching ratio is in the range 0.1--0.9.
Clearly, in the forbidden region, the $\lambda_{Hhh}$ cannot be
determined from resonant production.

\setcounter{equation}{0}
\section{Double Higgs-strahlung and Triple $h$ Production}
For small and moderate values of $\tan\beta$, the study of decays
of the heavy $CP$-even Higgs boson $H$ provides a means of determining
the triple-Higgs coupling $\lambda_{Hhh}$.
In order to extract the coupling $\lambda_{hhh}$, other processes
involving two-Higgs ($h$) final states must be considered.
The $Zhh$ final states, which can be produced in the non-resonant double 
Higgs-strahlung $e^+e^- \rightarrow Zhh$ , could provide one possible
opportunity, since it involves the coupling $\lambda_{hhh}$.
These non-resonant processes have also been investigated \cite{DHZ,OP98}.

\begin{figure}[htb]
\refstepcounter{figure}
\label{Fig:sig-Zll-2}
\addtocounter{figure}{-1}
\begin{center}
\setlength{\unitlength}{1cm}
\begin{picture}(15,8.5)
\put(-1.5,1)
{\mbox{\epsfysize=8.5cm\epsffile{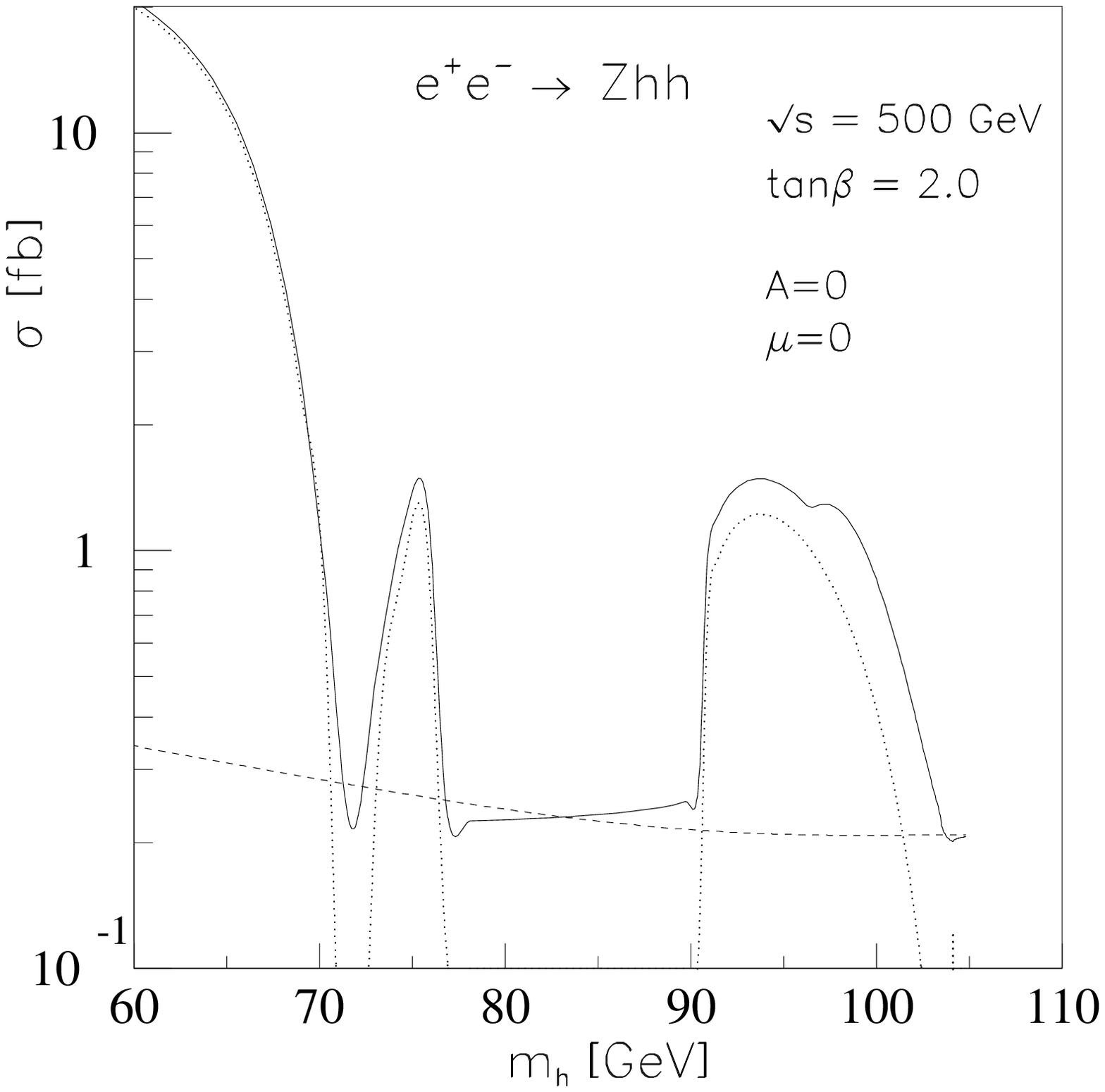}}
 \mbox{\epsfysize=8.5cm\epsffile{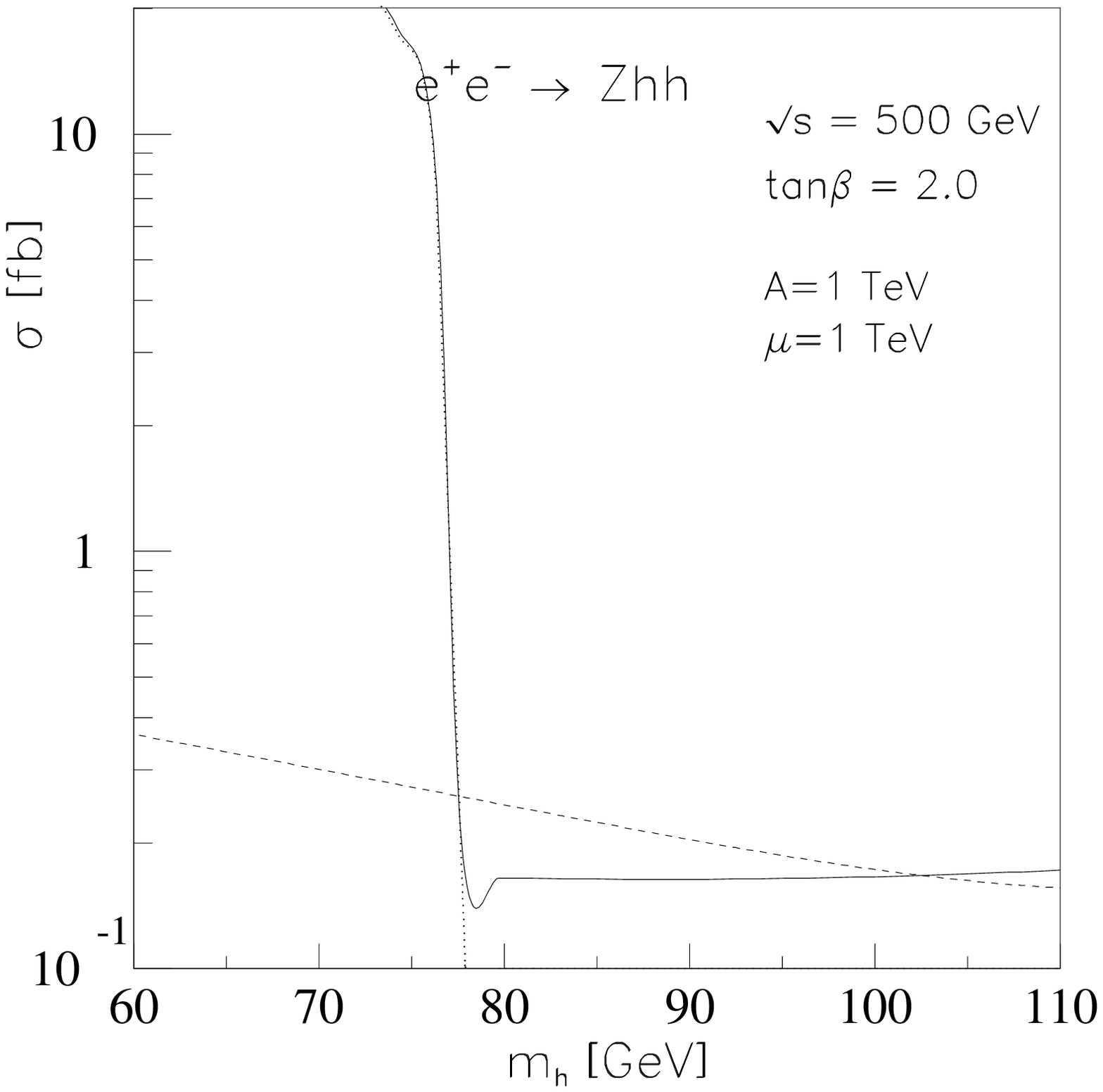}}}
\end{picture}
\vspace*{-8mm}
\caption{Cross section $\sigma(e^+e^-\to Zhh)$ as a function of $m_h$.
The dotted curve is the resonant production,
the dashed curve gives the decoupling limit.}
\end{center}
\end{figure}

We show in Fig.~\ref{Fig:sig-Zll-2} the $Zhh$ cross section, 
with $\tilde m = 1~\TeV$.
The structure around $m_h=70~\GeV$ (in the case of no mixing)
is due to the vanishing and near-vanishing of the trilinear coupling.

In the case of no mixing, there is a broad minimum from $m_h\simeq78$
to 90~GeV, followed by an enhancement around $m_h\sim90$--100~GeV.
This structure is due to the vanishing of the branching ratio 
for $H\to hh$, which is kinematically forbidden in the region 
$m_h\simeq78$--90~GeV, see Fig.~\ref{Fig:hole} (this coincides with 
the opening up of the channel $H\to WW$), 
followed by an increase of the trilinear couplings.
This particular structure depends considerably on the exact mass values
$m_H$ and $m_h$. Thus, it depends on details of the radiative corrections 
and on the mixing parameters $A$ and $\mu$. 

\setcounter{equation}{0}
\section{Fusion Mechanism for Multiple-$h$ Production}
A two-Higgs ($hh$) final state
in $e^+ e^-$ collisions can also result from the $WW$ fusion mechanism, 
which can either be a resonant process as in (\ref{Eq:res-Hhh}), 
or a non-resonant one like (\ref{Eq:WW-fusion}). 
Since the neutral-current couplings are smaller than 
the charged-current ones, 
the cross section for the $ZZ$ fusion mechanism in (\ref{Eq:res-Hhh}) 
and (\ref{Eq:WW-fusion}) is an order of magnitude smaller than the 
$WW$ fusion mechanism, and is here ignored.

The $WW$ fusion cross section for 
$e^+e^- \rightarrow H\bar\nu_e\nu_e$ can be written
as~\cite{DHKMZ} (see also \cite{OP98})
\begin{equation}
\sigma(e^+e^- \rightarrow H\bar\nu_e\nu_e) 
= \frac{G_F^3 m_W^4}{64 \sqrt 2\pi^3 }
\left[\int_{\mu_H}^1 dx\int_{x}^1 \frac{dy}
{\left[1 + (y-x)/\mu_W\right]^2}\ {\cal F}(x,y)\right]
\cos^2(\beta - \alpha).
\label{Eq:fusion-exact}
\end{equation}
This cross section is plotted in Fig.~\ref{Fig:sigma-500-1500} 
for the centre-of-mass energy $\sqrt s = 500$ GeV, 
and for $\tan\beta = 2.0$, as a function of $m_H$.  
The resonant fusion mechanism, which leads to $[hh]$ + [missing energy]
final states is competitive with the process
$e^+e^- \rightarrow HZ \rightarrow [hh]$ + [missing energy], 
particularly at high energies.
Since the dominant decay of $h$ will be into
$b\bar b$ pairs, the $H$-strahlung and the fusion mechanism will give 
rise to final states that will predominantly include four $b$-quarks.
On the other hand, the process $e^+e^- \rightarrow AH$ will give rise to
six $b$-quarks in the final state, since the $AH$ final state typically 
yields three-Higgs $h[hh]$ final states.

Besides the resonant $WW$ fusion mechanism for the multiple
production of $h$ bosons, there is also a non-resonant $WW$ 
fusion mechanism:
\begin{equation}
\label{Eq:WW-nonres}
e^+e^-\to\nu_e\bar\nu_e hh,
\end{equation}
through which the same final state of two $h$ bosons can be produced.
The cross section for this process, which arises
through $WW$ exchange as indicated in 
Fig.~\ref{Fig:Feynman-nonres-WW}, can be written in the
``effective $WW$ approximation'' as
\begin{equation}
\label{Eq:sigWW-nonres}
\sigma(e^+e^-\to\nu_e\bar\nu_e hh)
=\int_\tau^1\dd x\, \frac{\dd L}{\dd x}\, \hat\sigma\sup_{WW}(x),
\end{equation}
where $\tau=4m_h^2/s$. 
Here, the cross section is written as a $WW$ cross section,
at invariant energy squared $\hat s=xs$, 
folded with the $WW$ ``luminosity'' \cite{CDCG}:
\begin{equation}
\frac{\dd L(x)}{\dd x}
=\frac{G_{\rm F}^2m_W^4}{2}\,\left(\frac{v^2+a^2}{4\pi^2}\right)^2
\frac{1}{x}\biggl\{(1+x)\log\frac{1}{x}-2(1-x)\biggr\},
\end{equation}
where $v^2+a^2=2$.

The $WW$ cross section receives contributions from several amplitudes,
according to the diagrams (a)--(d) 
in Fig.~\ref{Fig:Feynman-nonres-WW}.
We have evaluated these contributions \cite{OP98}.

Our approach differs from that of DHZ in that we do not project out 
the longitudinal degrees of freedom of the intermediate $W$ bosons.
Instead, we follow the approach of Ref.~\cite{AMP}, where transverse
momenta are ignored everywhere except in the $W$ propagators.

\begin{figure}[htb]
\refstepcounter{figure}
\label{Fig:sig-WW-2}
\addtocounter{figure}{-1}
\begin{center}
\setlength{\unitlength}{1cm}
\begin{picture}(15,8.0)
\put(-1,1)
{\mbox{\epsfysize=8.5cm\epsffile{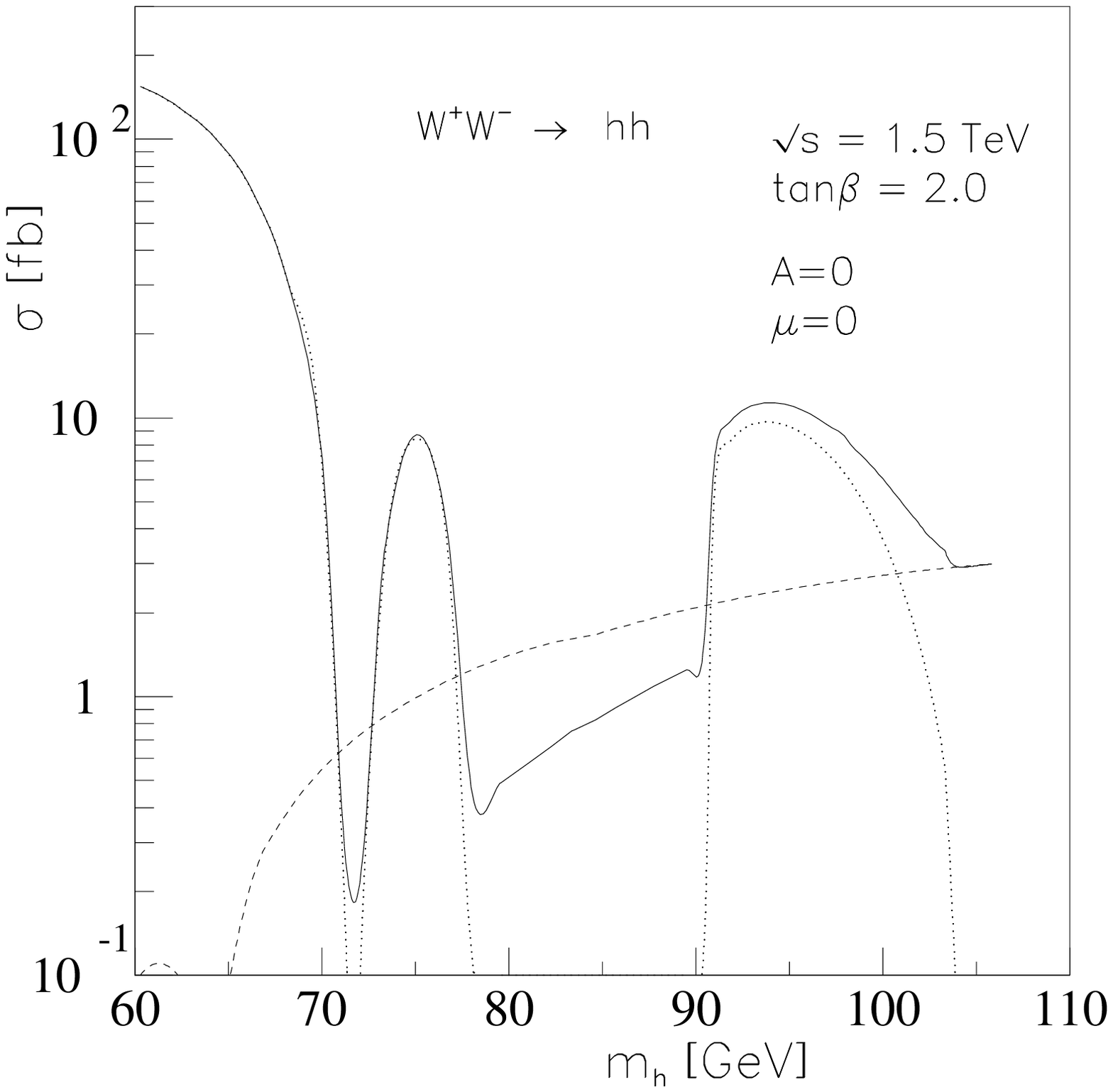}}
 \mbox{\epsfysize=8.5cm\epsffile{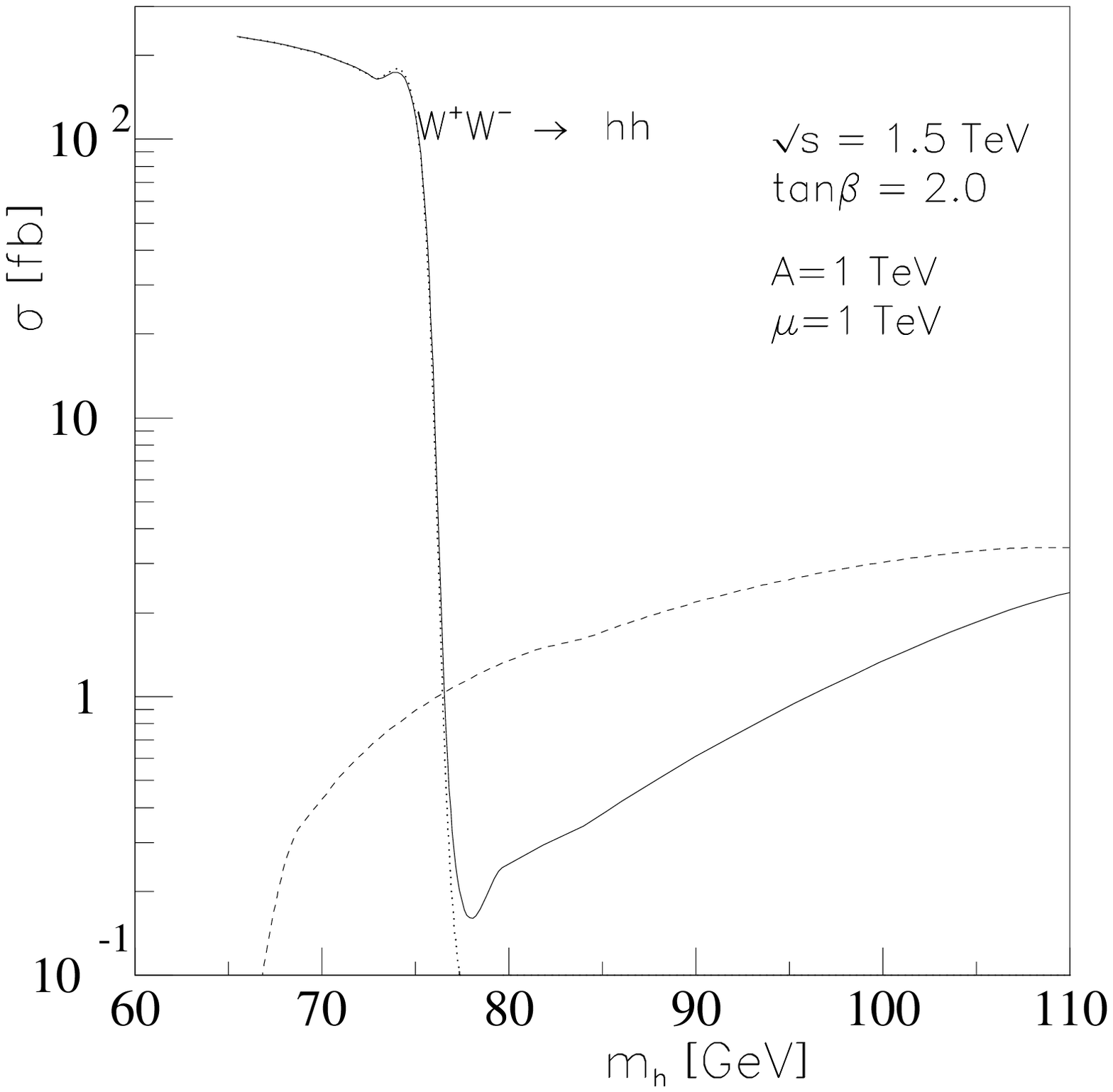}}}
\end{picture}
\vspace*{-8mm}
\caption{Cross section $\sigma(e^+e^-\to \nu_e\bar\nu_e hh)$ 
(via $WW$ fusion) as a function of $m_h$.
The dotted curve is the resonant production,
the dashed curve gives the decoupling limit.
}
\end{center}
\end{figure}

We show in Fig.~\ref{Fig:sig-WW-2} the $WW$ fusion cross section, 
at $\sqrt{s}=1.5~\TeV$,
as given by Eqs.~(\ref{Eq:fusion-exact}) and (\ref{Eq:sigWW-nonres}),
with $\tilde m = 1~\TeV$.
The structure is reminiscent of Fig.~\ref{Fig:sig-Zll-2},
and the reasons for this are same. Notice, however, that 
the scale is different.

\setcounter{equation}{0}
\section{Sensitivity to $\lambda_{Hhh}$ and $\lambda_{hhh}$}
In Fig.~\ref{Fig:sensi-500} we have indicated in the $m_A$--$\tan\beta$
plane the regions where $\lambda_{Hhh}$ and $\lambda_{hhh}$
might be measurable for $\sqrt{s}=500~\GeV$.
We identify regions according to the following criteria \cite{DHZ,OP98}:
\begin{itemize}
\item[(i)]
Regions where $\lambda_{Hhh}$ might become measurable are identified
as those where 
$\sigma(H)\times\mbox{BR}(H\to hh)> 0.1\mbox{ fb}$ (solid),
with the simultaneous requirement of 
$0.1 < \mbox{BR}(H\to hh) < 0.9$
[see Figs.~\ref{Fig:BR-H-A}--\ref{Fig:hole}].
In view of the recent, more optimistic, view on the
luminosity that might become available, 
we also give the corresponding contours for 0.05~fb (dashed) 
and 0.01~fb (dotted). 
\item[(ii)]
Regions where $\lambda_{hhh}$ might become measurable
are those where the {\it continuum} $WW\to hh$
cross section [Eq.~(\ref{Eq:sigWW-nonres})] is larger than 
0.1~fb (solid).
Also included are contours at 0.05 (dashed) and 0.01~fb (dotted).
\end{itemize}
Such regions are given for two cases of the mixing parameters
$A$ and $\mu$, as indicated.
We have excluded from the plots the region where $m_h<72.2~\GeV$,
according to the LEP lower bound \cite{ALEPH98}.
This corresponds to low values of $m_A$.

\begin{figure}[htb]
\refstepcounter{figure}
\label{Fig:sensi-500}
\addtocounter{figure}{-1}
\begin{center}
\setlength{\unitlength}{1cm}
\begin{picture}(15,8.0)
\put(-1.5,1)
{\mbox{\epsfysize=8.5cm\epsffile{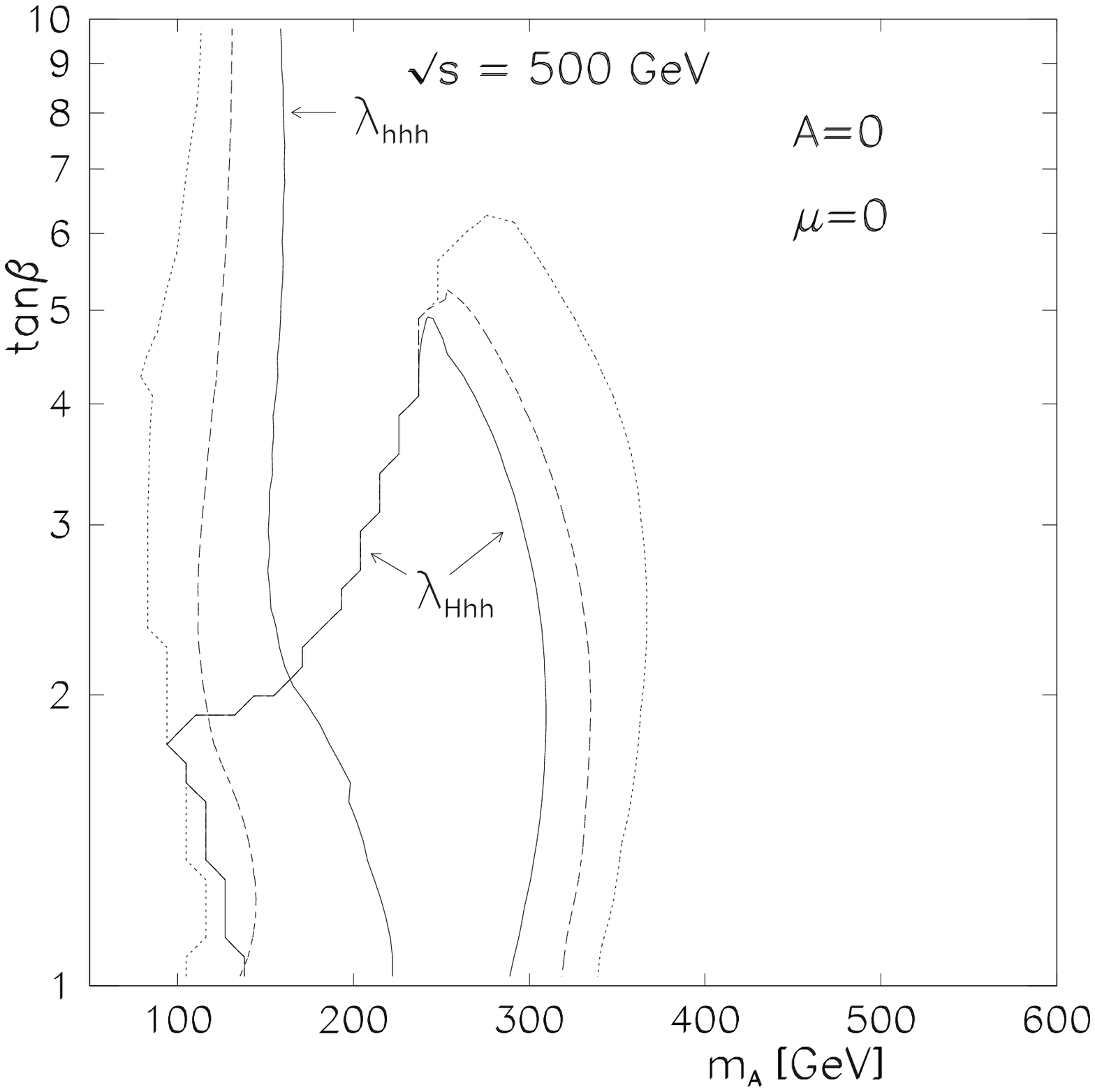}}
 \mbox{\epsfysize=8.5cm\epsffile{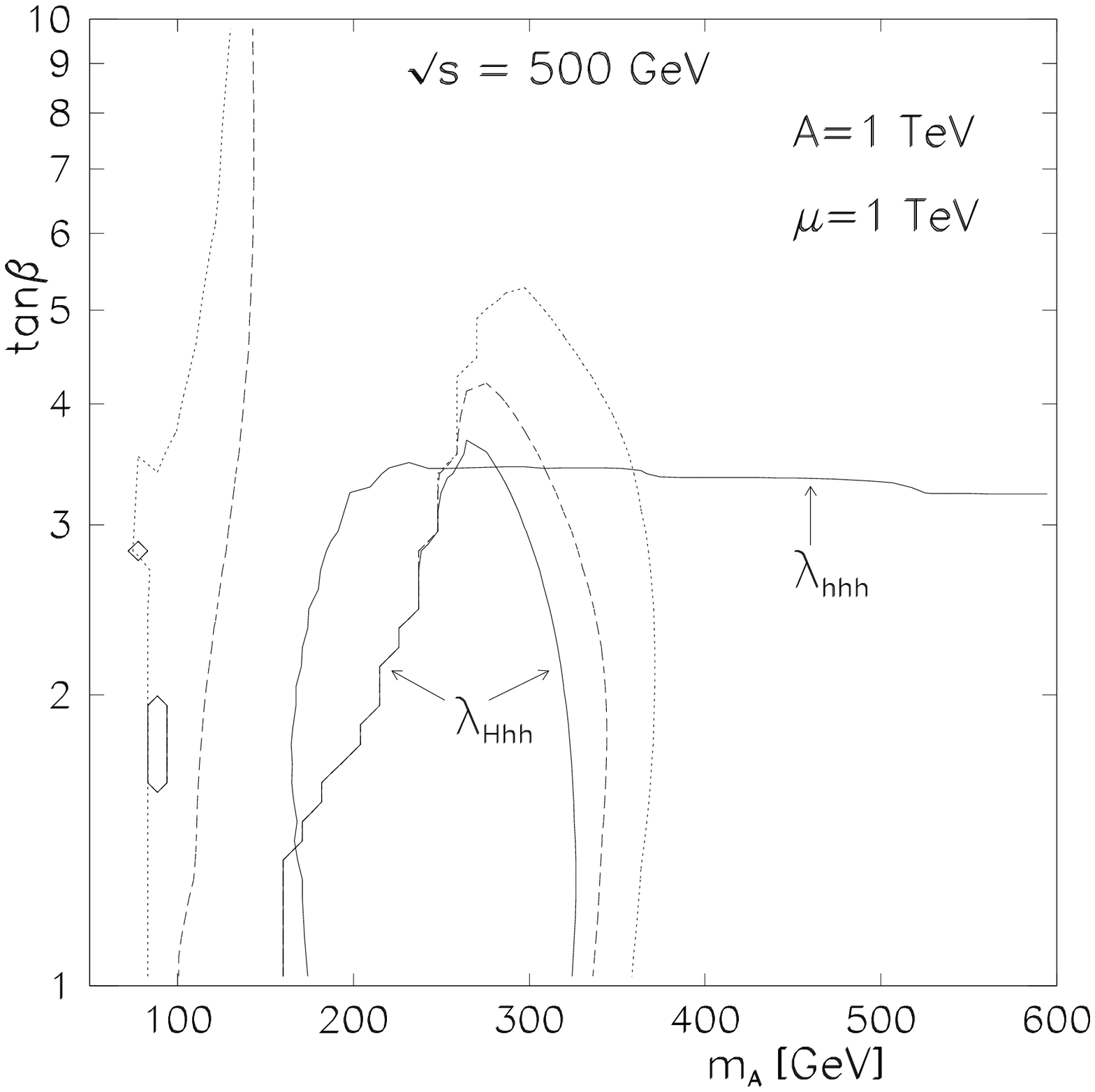}}}
\end{picture}
\vspace*{-8mm}
\caption{Regions where trilinear couplings $\lambda_{Hhh}$ and 
$\lambda_{hhh}$ might be measurable at $\sqrt{s}=500$~GeV.
Inside contours labelled $\lambda_{Hhh}$, 
$\sigma(H)\times\mbox{ BR}(H\to hh) > 0.1~\mbox{fb}$ (solid),
while $0.1<\mbox{BR}(H\to hh)<0.9$.
Inside (to the right or below) contour labelled $\lambda_{hhh}$,
the {\it continuum} $WW\to hh$ cross section exceeds 0.1~fb (solid).
Analogous contours are given for 0.05 (dashed) and 0.01~fb (dotted).
Two cases of mixing are considered, as indicated.
}
\end{center}
\end{figure}

With an integrated luminosity of 500~fb$^{-1}$,
the contours at 0.1~fb correspond to 50 events per year.
This will of course be reduced by efficiencies, but should indicate
the order of magnitude that can be reached.

At $\sqrt{s}=500~\GeV$, with a luminosity of 500~fb$^{-1}$ per year,
the trilinear coupling $\lambda_{Hhh}$ is accessible in a considerable
part of the $m_A$--$\tan\beta$ parameter space: at $m_A$ of the order
of 200--300~GeV and $\tan\beta$ up to the order of 5.
With increasing luminosity, the region extends somewhat 
to higher values of $m_A$.
The ``steep'' edge around $m_A\simeq200~\GeV$ (where increased luminosity
does not help) is determined by the vanishing of $\mbox{BR}(H\to hh)$,
see Fig.~\ref{Fig:hole}.

The coupling $\lambda_{hhh}$ is accessible in a much larger part
of this parameter space, but with a moderate luminosity,
``large'' values of $\tan\beta$ are accessible only if $A$ is small.

It should be stressed that the requirements discussed here
are necessary, but not sufficient conditions for the trilinear
couplings to be measurable. We also note that there might be 
sizable corrections to the $WW$ approximation, and that 
it would be desirable to incorporate the dominant two-loop
corrections to the trilinear couplings.
\setcounter{equation}{0}
\section{Conclusions}
We have presented the results of a detailed investigation \cite{OP98}
of the possibility of measuring
the MSSM trilinear couplings $\lambda_{Hhh}$ and $\lambda_{hhh}$
at an $e^+ e^-$ collider.
Where there is an overlap, we  
have confirmed the results of Ref.~\cite{DHZ}.
Our emphasis has been on taking into account
all the parameters of the MSSM Higgs sector.
We have studied the importance of mixing in the squark sector,
as induced by the trilinear coupling $A$ and the bilinear coupling $\mu$.

At moderate energies ($\sqrt{s}=500~\GeV$) the range in 
the $m_A$--$\tan\beta$ plane that is accessible for studying 
$\lambda_{Hhh}$ changes quantitatively for non-zero values of
the parameters $A$ and $\mu$.
As far as the coupling $\lambda_{hhh}$
is concerned, however, there is a qualitative change from the case of
no mixing in the squark sector.
If $A$ is large, then high luminosity is required to reach 
``high'' values of $\tan\beta$.
At higher energies ($\sqrt{s}=1.5~\TeV$), the mixing parameters
$A$ and $\mu$ change the accessible region of the
parameter space only in a quantitative manner.
\medskip

This research was supported by the Research Council of Norway,
and (PNP) by the University Grants Commission, India under project 
number 10-26/98(SR-I).


\end{document}